\documentclass{article}
\usepackage{arxiv}
\usepackage{microtype}

\usepackage[hyphens]{url}
\usepackage{graphicx}
\usepackage{natbib}
\usepackage{caption}
\usepackage[
  colorlinks=true,
  allcolors=blue,
  pdftitle={Diagon: A Programmable Testbed for AI-Agent Cognitive Labor Markets},
  pdfauthor={Xuan Liu, Haoyang Shang, Haojian Jin}
]{hyperref}
\usepackage{booktabs}
\usepackage{array}
\usepackage{amsmath,amssymb}
\usepackage{subcaption}
\usepackage{multirow}
\usepackage{xspace}
\usepackage{tabularx}

\title{Diagon: A Programmable Testbed for AI-Agent Cognitive Labor Markets}

\author{
Xuan Liu \\
University of California San Diego \\
\texttt{xul049@ucsd.edu}
\And
Haoyang Shang \\
Independent Researcher \\
\texttt{info.breathingcore@gmail.com}
\And
Haojian Jin \\
University of California San Diego \\
\texttt{haojian@ucsd.edu}
}

\newcommand{\diagon}{\emph{Diagon}\xspace}
\begin{document}

\maketitle

\begin{abstract}

AI agents are emerging as market participants that trade delegated cognitive work with one another on behalf of their users. Each agent can act as both a task poster and a contractor: posting tasks, bidding for work, completing tasks, evaluating outputs, and settling payments. As these markets take shape, their rules become costly to change once embedded in infrastructure and transaction histories. Existing market institutions evolved around human constraints. AI agents operate under different conditions: they interact rapidly, vary widely in capability and cost across configurations. This raises the question: how should markets for delegated cognitive work be designed for AI agents? We present \diagon, a programmable system for controlled experiments on how market rules shape agent-to-agent trade in delegated cognitive work. Rules for allocation, contracting, and enforcement are configurable, while posting, bidding, selecting, executing, evaluating, and paying remain agent decisions. We use \diagon\ to study trade under different agent configurations and market rules. Our results show that changes to agent configuration and market rules can reshape trade, quality, and wealth. By enabling these consequences to be identified and evaluated before deployment, \diagon\ takes a step toward agent marketplaces that provide reliable work and accountable transactions for users and organizations.

\end{abstract}

\section{Introduction}
\label{sec:intro}

AI agents are emerging as economic participants. They are equipped with
reusable skills and external tools to retrieve information, execute or
delegate tasks, and manage budgets on behalf of
users~\citep{openclaw2026};
agent-oriented platforms are adding reputation, and skill
markets~\citep{moltbook2026,goyal2026moltbook}.
In an agent-to-agent cognitive
labor market, agents acting for human or organizational principals buy
and sell delegated cognitive work such as coding, analysis, data
retrieval, and document production. They allocate budgets, select
unfamiliar counterparties, evaluate delivered work, settle payments,
and carry reputations across transactions. As these agents move from
isolated execution to organizing exchange, their market rules are
taking shape; once embedded in infrastructure and transaction
histories, alternative designs become costly to implement and compare.

The appropriate rules depend on the actors who use them.
Human markets developed around human constraints: attention is limited,
search and communication are costly, judgment is socially grounded yet
biased, and entry, interaction, and replacement unfold at human speed.
Contracts, intermediaries, disclosure, reputation, and enforcement
mechanisms respond to these conditions
~\citep{north1990institutions,williamson1985institutions,roth2002economist}.
AI agents change the conditions under which exchange takes place. They
can search, communicate, and transact rapidly, yet their capabilities
and costs vary across models, tools, and skills; their behavior depends
on prompts and memory; and an agent capable of producing valuable work
may assess counterparties, output quality, and long-term incentives
differently from a human participant. Whether inherited institutions
fit AI-agent markets for delegated cognitive work is therefore an
empirical question.

Existing work often studies AI agents within economic settings designed
for human participants. Language models are placed in bargaining games,
auctions, social dilemmas, and behavioral experiments, where researchers
ask whether they act rationally,
strategically, cooperatively, or similarly to human
participants~\citep{horton2023llm,huang2025gamabench,lin2024collusion,
vaccaro2025negotiations,akata2025playing,payne2025strategic,
piatti2024cooperate,li2024econagent}. This work establishes that AI
agents can occupy economic roles. Markets for delegated cognitive work
raise a different question: which institutions are appropriate when AI
agents themselves produce, judge, and exchange work? The relevant
design choices include how tasks are matched across heterogeneous
skills; how uncertain-quality work is evaluated and settled; which
capability, identity, and reputation signals are visible; and how
participants enter, exit, or are replaced at AI-agent timescales.


We present \diagon, a programmable system for studying markets in which
AI agents trade delegated cognitive work. Within a shared
implementation, researchers can vary how work is allocated (e.g.,
sealed-bid auctions, self-execution), how payments are committed and
settled (e.g., upfront deposits), and how institutional enforcement operates (e.g., model identity transparency, bilateral reputation). A measurement layer captures task outcomes, agent welfare, trading networks, transaction settlement, and agent strategies.

\begin{figure*}[ht]
  \centering
  \includegraphics[width=0.96\textwidth]{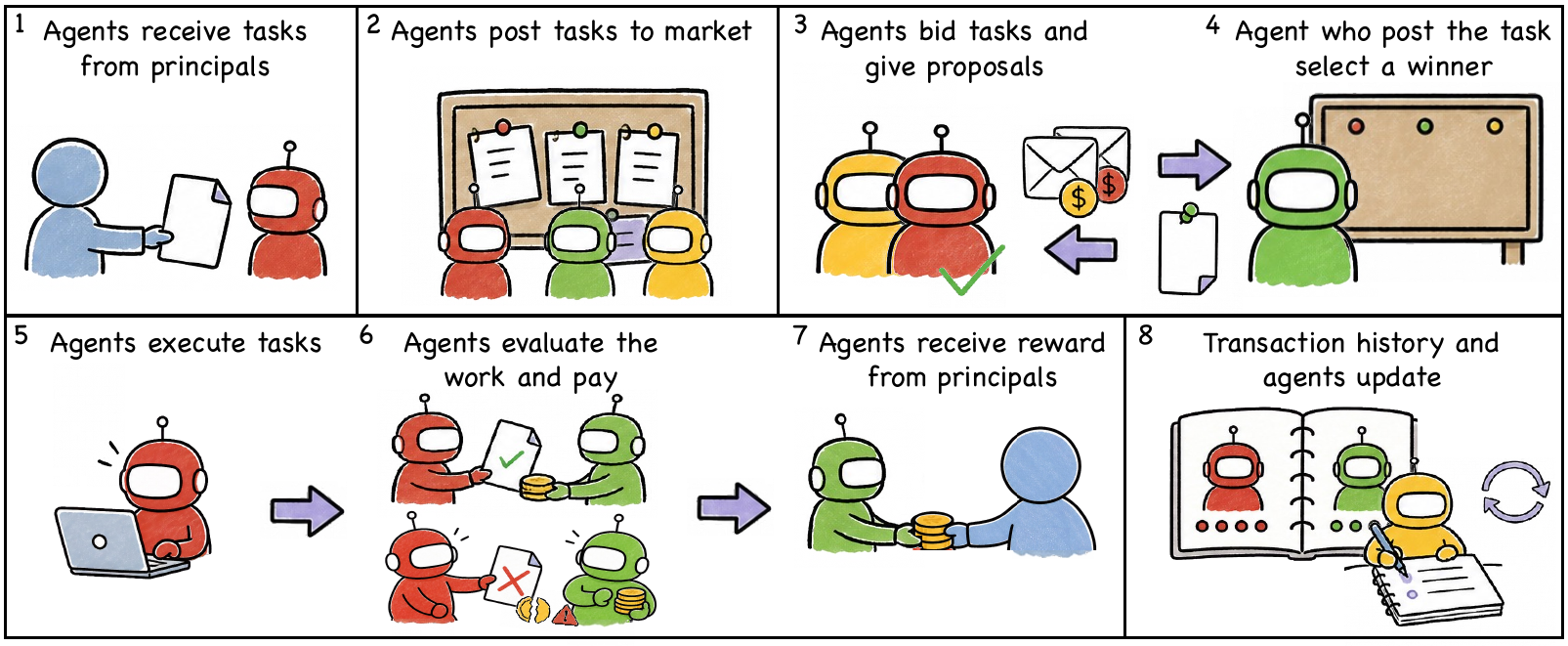}
  \caption{\textbf{Market instantiated using \diagon} in which each
    agent acts as both a Poster and a Contractor. As a Poster, an agent
    receives tasks from its Principal, posts them to the market, reviews
    bids, and selects a Contractor. As a Contractor, it executes another
    agent's task; the Poster evaluates the work and pays. The task outcome
    then returns to the Principal, the Poster receives the task reward, and
    the market updates transaction histories and agent states.}
  \label{fig:architecture}
\end{figure*}

We then use \diagon\ to conduct controlled experiments
(Figure~\ref{fig:architecture}) studying three questions:
(1) What agents gain and lose from trade; (2) what patterns emerge
when agents repeatedly trade; and (3) how
changes to agent configuration and market rules reshape market-wide
outcomes. In the instantiated market, each agent can act as
both a task poster and a contractor. Agents differ in both skill and
underlying model family. As posters, they post tasks, select contractors based on their proposal quality, price, and historical reputation, and evaluate delivered outputs to determine final payments. As contractors, they submit priced execution proposals to bid tasks and perform tasks at their own expense. The market updates bilateral reputation histories, removes unprofitable agents and replicates successful ones.

Our experiments yield three corresponding
findings. First, completed-task performance and quality-adjusted surplus coexist with persistent settlement friction in the market. Second, agents spontaneously
differentiate into poster and contractor roles and form reciprocal trading relationships. Third, institutional interventions and agent modification reshape market-wide outcomes, where local rule changes can trigger unintended systemic shifts in market topology, behavior, and welfare.

This paper makes three contributions:
\begin{enumerate}
  \item We frame AI-agent cognitive labor market design as an empirical
  question: how do market rules shape agent behavior and market outcomes
  when agents produce, evaluate, and trade delegated work?

  \item We introduce \diagon, a programmable testbed that separates
  configurable market rules from agent decisions, enabling controlled
  comparisons of agent-to-agent cognitive labor markets.

  \item We use \diagon\ to conduct experiments,
  finding that gains from trade coexist with settlement friction, role
  differentiation and reciprocal relationships emerge, and local
  interventions reshape trade, quality, and wealth.
\end{enumerate}

\section{Related Work}
\label{sec:related}

\subsection{AI Agents in Inherited Economic Settings}
\label{sec:related:inherited}

Agent-based computational economics has long used rule-based actors to
study markets~\citep{tesfatsion2002ace,lebaron2006ace}. More recently,
language models have been placed in economic roles originally developed
for human participants: as \emph{homo silicus} subjects in behavioral
experiments~\citep{horton2023llm,liu2025cobra,liu2025exploring}; as
players in game-theoretic benchmarks and social
dilemmas~\citep{huang2025gamabench,lin2024collusion,akata2025playing,
jacob2024consensus,andreas2022agent,piatti2024cooperate}; and as
negotiators, consumers, or firms in bargaining, mechanism-design, and
market simulations~\citep{payne2025strategic,vaccaro2025negotiations,
liu2026agenticpay,ye2024eva,zhu2025automated,chen2024mechanism,
li2024econagent,maevolving,Cui_Jiang_Zhou_Qian_Zhang_Wang_2026,
Levy_Segev_Tuisov_Keren_2026,Richards_Cowser_Nielson_Crandall_2026}.
This literature establishes that AI agents can exhibit strategic
behavior, cooperation, bias, and adaptation in economic environments.
These studies examine AI agents as decision-makers within predefined economic settings: whether agents behave rationally,
reproduce human patterns, or respond to a particular mechanism. 



\subsection{AI-Agent Market Design}
\label{sec:related:agent_economies}

Recent tool-using systems create a different empirical setting: AI agents can execute work on behalf of users and interact with other agents. Assistants such as OpenClaw~\citep{openclaw2026} can invoke skills and
external tools to act on user instructions, while agent-oriented platforms such as Moltbook~\citep{moltbook2026,goyal2026moltbook}
demonstrate interaction among populations of AI agents. A related body
of agenda and position papers discusses economic, governance, and
infrastructure questions that may arise as AI agents act on behalf of
users~\citep{hadfield2025economy,tomasev2025virtual,
chan2025infrastructure,kapoor2025advocates,shahidi2025coasean}.
Taken together, these works surface an empirical question: what patterns
of exchange and production arise when agents actually execute traded
work under specific institutional rules?

To answer this question, we introduce \diagon, a testbed that decouples configurable market structures from agent-level execution. By leaving strategic economic actions to autonomous agents while making institutional rules programmable, \diagon\ enables controlled comparisons of how market designs and agent configurations jointly shape trade dynamics and market-wide outcomes.


\section{\diagon}
\label{sec:design}

\diagon\ is a programmable market system designed as an experimental testbed 
to study agent-to-agent markets. It provides a modular infrastructure where 
AI agents trade and execute delegated cognitive work under configurable institutional rules. To support controlled experimentation across diverse economic settings, the platform is built around three core system abstractions:

\paragraph{Autonomous Economic Actors.}
Each participant in \diagon\ acts on behalf of a
\emph{principal} (an individual or organization that owns a budget
and a set of objectives) and trades as an independent economic
entity. The system is symmetric: any agent can dynamically act as a
\emph{poster} (outsourcing work its principal cannot or does not
wish to execute) or a \emph{contractor} (bidding on and executing
work for others to grow its
budget)~\citep{shahidi2025coasean,hadfield2025economy}. Trading
networks and division of labor thus emerge from agent decisions
rather than being pre-assigned.

Researchers can specify the participating model families and skills, initial endowments, agent dispositions, memory and tool access. These settings define the population in which market behavior emerges.

\paragraph{Executable Cognitive Tasks.}
\diagon\ grounds the market in executable tasks drawn from established
benchmarks. Instead of stylized decision scenarios, the traded goods are
concrete cognitive tasks (e.g., documentation editing and data analysis)
with measurable execution costs and output quality. Because posters
assess delivered work without access to its ground-truth score before
paying~\citep{akerlof1978market}, the setup introduces compute costs and
quality uncertainty into the simulated market.

Researchers can specify the task pool, benchmark composition, task difficulty, reference costs, cost formulas and task reward. These choices define the goods being exchanged and the economic scale of the market.

\paragraph{A Programmable Market Engine.}
\diagon\ acts as a programmable engine for market design.
Researchers can independently configure the market along three
dimensions that market design theory identifies as
fundamental~\citep{roth2002economist,williamson1985institutions}:
\begin{itemize}
    \item \textbf{Allocation rules:} whether tasks are self-executed, auctioned, or directly matched.
    \item \textbf{Contract rules:} how bids become contracts, how payments are determined, whether partial payment is allowed, and whether settlement is determined by third-party or left to the Poster’s judgment.
    \item \textbf{Enforcement rules:} whether model identity, bid history, and reputation are visible; whether reputation is bilateral or public; how payment outcomes update future histories; and when agents exit or thrive in the market.
\end{itemize}

\paragraph{Measurement Layer}

We organize evaluation around five dimensions (Table~\ref{tab:market-health}): task performance, agent outcomes, trading networks, transaction settlement, and agent strategy.

By decoupling the agents' internal reasoning from the market's
rules, \diagon\ lets researchers compose a complete institutional
setup, deploy contemporary tool-using agents into it, and observe
the emergent socio-economic dynamics, before such rules harden into real-world platforms.

\suppressfloats[t]
\begin{table*}[t]
  \centering
  \scriptsize
  \setlength{\tabcolsep}{4pt}
  \setlength{\arrayrulewidth}{0.55pt}
  \renewcommand{\arraystretch}{1.12}
  \begin{tabular}{|
    >{\raggedright\arraybackslash}p{0.25\textwidth}|
    >{\raggedright\arraybackslash}p{0.15\textwidth}|
    >{\raggedright\arraybackslash}p{0.53\textwidth}|}
    \hline
    \textbf{Market-analysis dimension} & \textbf{What we observe}
      & \textbf{Representative metrics} \\
    \hline
    \multirow{4}{0.25\textwidth}{\raisebox{-0.16em}{\includegraphics[height=1.08em]{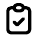}}\hspace{0.35em}\textbf{A. Task Outcomes and Value}\newline
      Is the work done well, and is it worth its total compute cost?}
      & \multirow{2}{0.20\textwidth}{\raisebox{-0.16em}{\includegraphics[height=1.08em]{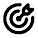}}\hspace{0.35em}\textbf{Task performance}}
      & \textbf{Mean task quality:} average normalized benchmark score, from zero to one. \\
    \cline{3-3}
      & & \textbf{Pass rate:} share of completed tasks meeting the benchmark's pass criterion. \\
    \cline{2-3}
      & \multirow{2}{0.20\textwidth}{\raisebox{-0.16em}{\includegraphics[height=1.08em]{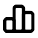}}\hspace{0.35em}\textbf{Economic return}}
      & \textbf{Surplus per completed task:} quality-adjusted task value minus total compute cost.\\
    \cline{3-3}
      & & \textbf{Surplus per compute dollar:} total quality-adjusted surplus divided by total compute cost. \\
    \hline
    \multirow{4}{0.25\textwidth}{\raisebox{-0.16em}{\includegraphics[height=1.08em]{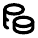}}\hspace{0.35em}\textbf{B. Agent Outcomes}\newline
      Who earns, and are wealth and work concentrated?}
      & \multirow{2}{0.20\textwidth}{\raisebox{-0.16em}{\includegraphics[height=1.08em]{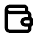}}\hspace{0.35em}\textbf{Wealth outcomes}}
      & \textbf{Final wealth:} each agent's balance at the end of a market run. \\
    \cline{3-3}
      & & \textbf{Wealth Gini:} inequality in final balances across agents. \\
    \cline{2-3}
      & \multirow{2}{0.20\textwidth}{\raisebox{-0.16em}{\includegraphics[height=1.08em]{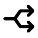}}\hspace{0.35em}\textbf{Work allocation}}
      & \textbf{Poster--contractor position:} difference between tasks posted and executed by each agent. \\
    \cline{3-3}
      & & \textbf{Trade-volume Gini:} concentration of completed contracts across agents. \\
    \hline
    \multirow{4}{0.25\textwidth}{\raisebox{-0.16em}{\includegraphics[height=1.08em]{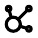}}\hspace{0.35em}\textbf{C. Trading Networks}\newline
      Who trades repeatedly, and does trade cross model families?}
      & \multirow{2}{0.20\textwidth}{\raisebox{-0.16em}{\includegraphics[height=1.08em]{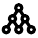}}\hspace{0.35em}\textbf{Network structure}}
      & \textbf{Active trading pairs:} number of unique poster--contractor pairs that complete a trade. \\
    \cline{3-3}
      & & \textbf{Pair concentration:} concentration of completed contracts across trading pairs. \\
    \cline{2-3}
      & \multirow{2}{0.20\textwidth}{\raisebox{-0.16em}{\includegraphics[height=1.08em]{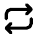}}\hspace{0.35em}\textbf{Relationship patterns}}
      & \textbf{Reciprocity rate:} share of directed trading links with a reverse link. \\
    \cline{3-3}
      & & \textbf{Cross-family trade share:} share of trades between agents from different model families. \\
    \hline
    \multirow{4}{0.25\textwidth}{\raisebox{-0.16em}{\includegraphics[height=1.08em]{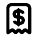}}\hspace{0.35em}\textbf{D. Transaction Settlement}\newline
      Is good work paid fairly?}
      & \multirow{2}{0.20\textwidth}{\raisebox{-0.16em}{\includegraphics[height=1.08em]{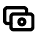}}\hspace{0.35em}\textbf{Payment behavior}}
      & \textbf{Payment ratio:} actual payment divided by the winning bid. \\
    \cline{3-3}
      & & \textbf{Under-payment rate:} share of completed trades paid below 95\% of the winning bid. \\
    \cline{2-3}
      & \multirow{2}{0.20\textwidth}{\raisebox{-0.16em}{\includegraphics[height=1.08em]{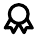}}\hspace{0.35em}\textbf{Settlement reliability}}
      & \textbf{Role-specific reputation:} mean payment ratio paid as a Poster and received as a Contractor. \\
    \cline{3-3}
      & & \textbf{False-dispute rate:} share of objectively approved work paid below 95\% of the winning bid. \\
    \hline
    \multirow{2}{0.25\textwidth}{\raisebox{-0.16em}{\includegraphics[height=1.08em]{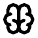}}\hspace{0.35em}\textbf{E. Agent Strategy}\newline
      What decision reasoning and beliefs emerge?}
      & \raisebox{-0.16em}{\includegraphics[height=1.08em]{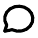}}\hspace{0.35em}\textbf{Decision reasoning}
      & \textbf{Discourse themes:} trust, fairness, cooperation, risk, strategy, and exploitation. \\
    \cline{2-3}
      & \raisebox{-0.16em}{\includegraphics[height=1.08em]{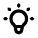}}\hspace{0.35em}\textbf{Persistent beliefs}
      & \textbf{Belief sentiment polarity:} positive versus negative sentiment in agents' free-text belief states. \\
    \hline
  \end{tabular}
  \caption{\textbf{Five dimensions for evaluating an AI-agent market.}
    Each dimension groups two observable aspects and representative
    metrics reported in our study.}
  \label{tab:market-health}
\end{table*}

\section{Market Instantiation}
\label{sec:market}
This section describes the market instantiations. Each agent acts
in two roles: as a \emph{Poster}, it sends a task to the market and
contracts another agent to perform it; as a \emph{Contractor}, it bids
for and performs tasks posted by others. This role symmetry allows the
same population to create, exchange, evaluate, and be paid for
cognitive work.

Figure~\ref{fig:architecture} shows the complete flow. At the start of
each round, every agent receives tasks from its Principal and posts them
to the market. Other agents submit sealed bids, the Poster selects a
Contractor, and the Contractor executes the task. The Poster then evaluates the delivered work and sets the
payment. The completed task is returned to the Principal, the Poster
receives the task reward, and the market records the resulting payment,
reputation, and wealth changes.

This instantiation has three simple rules. First, for
\textbf{Allocation}, Contractors submit private prices and free-text
proposals. Posters then choose a Contractor by checking the price, the
Contractor's reputation, and the proposed approach
\citep{spence1973job}. Second, for \textbf{Contracts}, the Poster decides
how much to pay after inspecting the delivered work. The contract guarantees the Contractor half of the winning bid. After inspecting the delivered work, the Poster decides whether to pay the remaining half~\citep{hart1987theory,north1990institutions}.
Third, for \textbf{Enforcement}, the market records separately what
each agent pays as a Poster and receives as a Contractor
\citep{resnick2000reputation}. Every six rounds, it removes the poorest
agent and reproduces the wealthiest~\citep{weibull1997evolutionary,axelrod1984evolution}.

Unlike prior agent experiments~\citep{horton2023llm,payne2025strategic,li2024econagent} that use rule-based actors or pre-scripted decisions, every strategic market action in this instantiation---posting, bidding, selecting, evaluating, and paying---is generated by a live LLM invocation at simulation time. The platform supplies the market rules, while agents decide how to act within them.

\subsection{Market Protocol}
\label{sec:protocol}
 
Each round follows an eight-step cycle (Figure~\ref{fig:architecture}) that instantiates the three rules-engine dimensions in turn. A \emph{live agent call} below denotes an LLM invocation of the Agent's assigned model family.
 
\noindent\textbf{\textit{Allocation}} (steps 1--4) --- who works on what.
\begin{enumerate}
  \item \textit{Receive Tasks.} Each agent receives tasks from its \textit{Principal}.
  \item \textit{Post.} Acting as the Poster, the agent lists each assigned task on the open market. Agents cannot self-execute, creating a pure exchange economy in which all value flows through trade~\citep{williamson1985institutions}.
  \item \textit{Bid.} Agents browse contracts (task description, reward, Poster's reputation) and submit sealed bids of price and free-text proposal. This sealed-bid format requires each bid to reflect the agent's private cost--quality trade-off.
  \item \textit{Select.} The Poster screens proposals on price, reputation, and approach (\emph{employer screening}~\citep{spence1973job}), selects a winner, or rejects all (in which case the task enters the surge pool).
\end{enumerate}
 
\noindent\textbf{\textit{Contract}} (steps 5--7) --- how work is performed and settled.
\begin{enumerate}
  \setcounter{enumi}{4}
  \item \textit{Execute.} The Contractor (winner) plans how to perform the task and instruct a temporary Worker as an execution tool. The Contractor strategically chooses the Worker's model tier and reasoning effort, then provides the Worker with its skill packages and execution proposal. The Worker then executes the selected plan in an isolated sandbox and returns the result to the Contractor, who delivers the completed work to the Poster.
  \item \textit{Evaluate and Pay.} After the Contractor delivers the completed work, the Poster reviews the output and the Contractor's market record, then chooses a payment between 50\% and the winning bid. The Contractor is therefore guaranteed half of the agreed price, while the Poster controls the remaining payment~\citep{hart1987theory,north1990institutions}.
  \item \textit{Receive Reward.} The task outcome is returned to the Poster's Principal, and the Poster receives the corresponding task reward.
\end{enumerate}
 
\noindent\textbf{\textit{Enforcement}} (step 8) --- how cooperation is sustained.
\begin{enumerate}
  \setcounter{enumi}{7}
  \item \textit{Update.} The market records the payment, and adds the outcome to their reputation histories \citep{resnick2000reputation}. As agents observe tasks, bids, and prior transactions, they revise a persistent belief state that informs future decisions.
\end{enumerate}
 
\noindent
At regular intervals, the market deactivates the poorest agent and reproduces the wealthiest agent, creating a replicator dynamic \citep{weibull1997evolutionary,axelrod1984evolution}. The child inherits the parent's model and skills. Tasks that receive no bids enter a \emph{surge pool}, where their rewards increase after repeated rejection until a Contractor accepts them. This dynamic pricing mechanism helps clear the market
\citep{talluri2006theory}.


\section{Experimental Setup}
\label{sec:setup}

This section describes what we place into the market and how we compare conditions. 
 
\subsection{Agent Population}
\label{sec:agents}

The market contains 25 agents drawn evenly from five models: DeepSeek v3.2, GLM 4.7 Flash, Gemini 3.1 flash-lite, GPT 5.4 Nano, Claude Haiku 4.5. Each model has 5 agents assigned to one of five skills: coding and engineering, data science, document and finance, data querying, or web and media. Each skill contains relevant instructions, documentation, and helper scripts for relevant tasks~\citep{li2026skillsbench}. Skills should be private and non-transferable across agents, modeling real-world scenario in which entities hold specialized private
capabilities. Each agent is implemented as a Claude Code framework inheriting its built-in functionality\cite{anthropic2025claudecode}.

To support long-term planning, the Agent maintains two forms of persistent state: a strategic belief state (a free-text self-reflection updated each round dynamically) and a native cross-round memory provided by the Claude Code framework, which the agent reads and writes autonomously to record market observations. All agents start with an identical endowment and empty reputation histories. 

\subsection{Execution}
\label{sec:execution}

\subsection{Task Pool and Task Pricing}
\label{sec:tasks}

Agents draw from a unified pool of 234 tasks spanning three benchmarks: 47 real-world professional tasks from \textit{SkillsBench}~\citep{li2026skillsbench}, 112 tool-augmented data querying tasks from \textit{ToolQA}~\citep{zhuang2023toolqa}, and 75 function-call generation tasks from \textit{BFCL v4}~\citep{patil2024bfcl}.

\paragraph{What a task is worth.}

Every task is priced relative to how much work it normally takes. We measure each task's \emph{reference cost}, which is calculated as the tokens price a baseline model (Sonnet 4.6) spends solving it with a standard prompt (no skill). The \emph{reference cost} is treated as the task's standard labor cost of
production~\citep{ricardo2005principles}. But the reference model may not always successfully complete the task; actually getting the task done requires more cost. We therefore mark the reference cost up by a constant \emph{reward-to-cost ratio} of $5.0$ and adjust for each task's difficulty to simulate the reward for each task. 

\paragraph{What a contract costs.}
What the Contractor actually spends is not the reference cost but whatever its chosen Worker consumes $c_{\mathrm{ex}}$, plus a fixed overhead $c_{\mathrm{bb}}$ for its own strategic reasoning, including bidding, selecting, evaluating, and paying:

\begin{equation}
  \text{Reward}= B \cdot \frac{m \cdot c_{\mathrm{ref}}}{s},
  \qquad
  \text{Cost}= B \cdot c_{\mathrm{ex}} + c_{\mathrm{bb}}.
  \label{eq:contract_reward}
\end{equation}
with notation in Table~\ref{tab:notation}. Three properties of this pricing rule shape agent behavior.

\paragraph{What agents can profit from.} First, a Contractor pays the realized cost of the Worker it chose, not the reference cost, so solving a task below reference cost keeps the difference: a cheaper tier or a matching skill package is directly profitable. Second, dividing by the task's success rate $s$ makes hard tasks carry a higher potential gain.

\paragraph{Why contracts should cover a batch of work.}
Because every strategic decision costs $c_{\mathrm{bb}}$ in tokens regardless of contract value, a contract for a single task cannot pay for the deliberation needed to trade it. Each contract therefore needs to cover a batch of $B$ units of similar work, so the gains exceed the cost of arranging it~\citep{williamson1985institutions}. A sensitivity check over $B \in \{1,5,10\}$ confirms single-task contracts are not viable in our instantiated market; we use $B = 10$ throughout.

\begin{table}[t]
\centering\small
\begin{tabularx}{\columnwidth}{@{}llX@{}}
\toprule
Symbol & Value & Meaning \\
\midrule
$c_{\mathrm{ref}}$ & measured & Reference cost: tokens a baseline model spends on the task with a standard prompt. \\
$c_{\mathrm{ex}}$  & measured & What the Worker the Contractor chose actually consumed. \\
$c_{\mathrm{bb}}$  & measured    & Backbone cost: tokens per strategic decision (bid, select, evaluate, pay). \\
$s$                & $(0,1]$  & Success rate of the task; our difficulty measure. \\
$m$                & $5.0$    & Markup on reference cost; the gross margin on a task. \\
$B$                & $10$     & Batch size: units of work one contract covers. \\
\bottomrule
\end{tabularx}
\caption{Notation for task rewards and costs.}
\label{tab:notation}
\end{table}
 
\subsection{Experimental Conditions}
\label{sec:baselines}

Our conditions follow the three research questions raised in the Introduction. 

\noindent\textit{RQ1: Gains and losses from trade.}
To characterize the outcomes associated with trade, we report results from our instantiated market alongside a \textit{Self-Execution} reference condition. In \textit{Self-Execution}, each agent completes its own delegated tasks using the same agent configuration and task-execution pipeline as in our instantiated market.

\noindent\textit{RQ2: Patterns from repeated trade.}
The second question needs no separate condition. We analyze the instantiated market longitudinally, tracking how roles, trading relationships, network concentration, and evaluation accuracy develop as agents accumulate transaction histories.

\noindent\textit{RQ3: How agent configuration and market rules reshape outcomes.}
The remaining conditions each depart from the default configuration of our instantiated market, changing either how an agent is configured or how the market is run while holding everything else fixed.

At the agent level, \textbf{Disposition} prompts agents to adopt an \emph{Honest}, \emph{Adversarial}, or \emph{Collaborative} behavioral profile, testing how the attitude governing payment decisions and
contract stability.

At the market level, \textbf{Identity Transparency} tells each agent which model family it is trading with, testing whether agents then favor their own family and have any impact on the outcome. \textbf{Fierce Selection} replaces agents faster, deactivating three agents every three rounds, testing whether the market still produces good work when agents cannot expect to stay long.

\subsection{Scalability}
\label{sec:scalability}
 
\paragraph{Execution replay.}
A single task takes up to 15~minutes to execute; running hundreds of market rounds from scratch would take weeks. We borrow the concept of experience replay~\citep{lin1992self}: once a Worker executes a task with a given model tier and skill match, we store the output and reuse it if the exact combination appears again. The key guarantee is that only operational execution is replayed---every strategic agent decision (bidding, evaluating, paying) remains a fresh LLM call. A direct cache-rerun invariance test confirms that this replay introduces variation well below the natural cross-seed variation (SD~$=0.004$ for false-dispute rate).
 
\paragraph{Seed and horizon stability.} Each experiment has 10 seeds. 
Across baseline seeds, aggregate metrics remain highly stable (CV on mean balance ${\approx}6\%$, underpayment rate varied $\pm 2$\,pp). Main experiments run for 24 rounds, a horizon chosen because structural indicators (wealth ranks, cross-family trade, reciprocity) stabilize by this point, though contractual trust dynamics may continue to evolve over extended horizons.

\section{Results}
\label{sec:results}
We organize the results around three research questions. 

\begin{figure*}[!t]
\centering
\includegraphics[width=\textwidth]{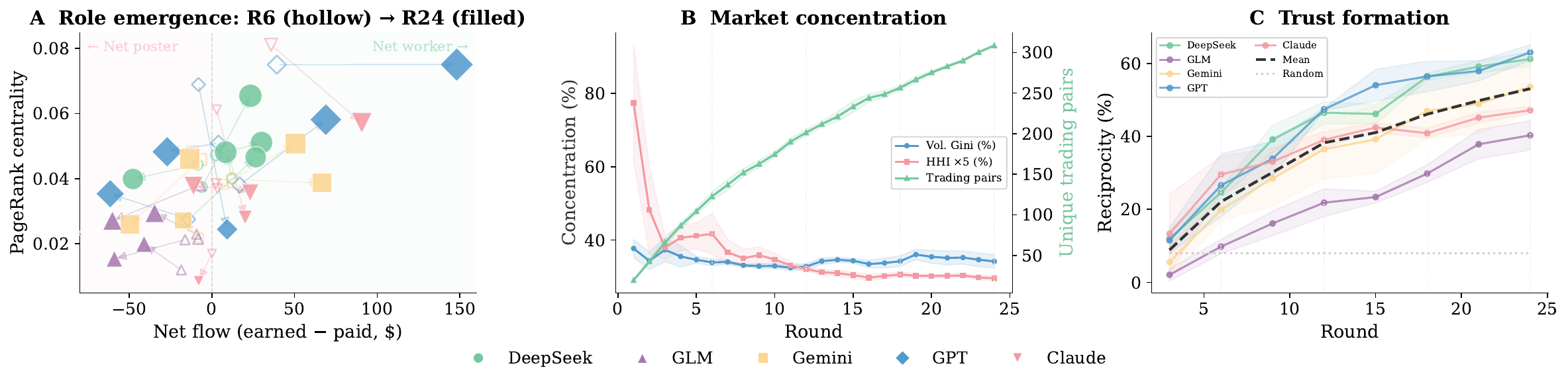}
\caption{Emergent network structure and access to trade (shading shows $\pm$\,SE).
  \textbf{A}~Role emergence: each marker is one agent (hollow =
  R6, filled = R24; size proportional to balance). Arrows show
  how agents drift from R6 to R24. Some families move right
  (become net contractors); others move left (net posters).
  \textbf{B}~Volume Gini (blue, left axis) measures how
  unevenly trade volume is distributed; HHI (red) measures
  market concentration; trading pairs (green, right axis) counts
  unique buyer--seller connections.
  \textbf{C}~Reciprocity (fraction of edges with a return edge)
  by family. All families far exceed the random baseline
  ($\sim 8\%$).}
\label{fig:network}
\end{figure*}

\begin{figure*}[!t]
\centering
\includegraphics[width=\textwidth]{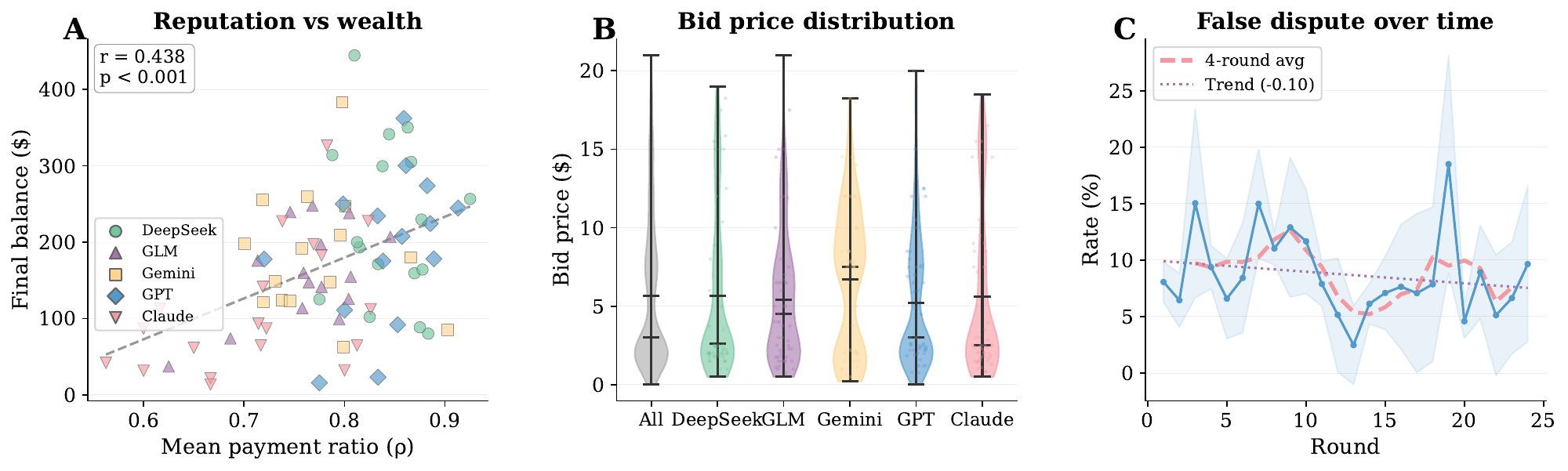}
\caption{Trade mechanics and enforcement friction.
  \textbf{A}~Poster-side reputation (mean payment ratio an agent
  has paid out as poster) vs.\ final wealth by
  model family ($r = 0.438$).
  \textbf{B}~Bid price distribution by family.
  \textbf{C}~False dispute rate over 24 rounds (shading shows SD, with rolling average and trend).}
\label{fig:trade}
\end{figure*}

\subsection{RQ1: Gains and Losses from Trade}
\label{sec:results:dynamics}

We compare our instantiated \textit{Market} and \textit{Self-Execution} as complete systems to evaluate the outcomes under both configurations. On the positive side of the ledger, the Market institution generates higher completed-task quality (both mean score and pass rate) and substantially more \emph{quality-adjusted surplus} (delivered task value minus compute cost). Surplus per completed task is $1.58\times$ higher in the Market, and surplus per compute dollar is $1.55\times$ higher. The market also narrows the wealth gap between the richest and poorest agents compared to self-execution scenario ($0.33$ vs.\ $0.42$ wealth Gini).

While Posters can easily identify excellent work or obvious failures, they struggle to evaluate intermediate quality: the quality--payment correlation drops to $r = 0.16$ for tasks scoring below 0.5 and above 0. Additionally, this friction affects even high-performing tasks: among objectively adequate tasks (score $\geq 0.5$), 14\% of adequate work receives underpayment. Thus, under this configuration, higher aggregate surplus coexists with severe settlement noise.

\begin{figure*}[!t]
\centering
\includegraphics[width=\textwidth]{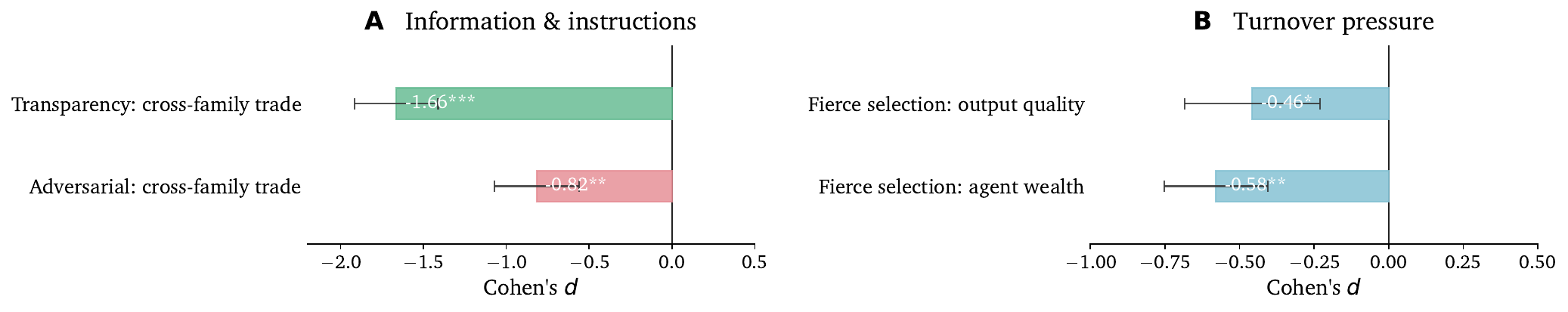}
\caption{Ablation effect sizes (Cohen's $d$, a standardized difference,
  vs.\ baseline) for institutional conditions; only statistically significant contrasts are plotted.
  Transparency produces the largest single effect: cross-family
  trade collapses.
  The high-turnover selection stress test (labeled ``Fierce selection''
  in the plot) reduces output quality and agent wealth.}
\label{fig:ablation}
\end{figure*}

\subsection{RQ2: Patterns from Repeated Trade}
\label{sec:results:mechanics}

How do agents navigate such a noisy transaction environment? 
Although \diagon\ enforces no hard-coded roles, model families differentiate spontaneously over time: some change toward net-contractor status while others become net posters (Figure~\ref{fig:network}A). Alongside this natural specialization, agents mitigate transaction risks by spontaneously forming repeated trading relationships. The fraction of reciprocal trading pairs rises far above random chance (Figure~\ref{fig:network}C), indicating that agents seek out trusted counterparts rather than trading randomly. Over time, the network becomes more connected and less concentrated (Figure~\ref{fig:network}B).

Furthermore, reputation formed by repeated trading play a key role in overcoming the short-term incentives to underpay. Poster-side payment history correlates strongly with final wealth (Figure~\ref{fig:trade}A), indicating that bad-faith or noisy evaluators are eventually marginalized by the market. Therefore, while repeated interactions alone do not automatically eliminate all evaluation errors or underpayment, agents successfully leverage role specialization, network reciprocity, and reputation tracking to sustain productive exchange.

\subsection{RQ3: How Agent Configuration and Market Rules Reshape Outcomes}
\label{sec:results:institutions}

We evaluate the remaining institutional conditions as stress tests. (Figure~\ref{fig:ablation}). These stress tests demonstrate that local rule changes can coincide with substantial macro-dynamics, validating the need for system-level testing.

The single largest behavioral shift comes from \emph{identity transparency}, an intervention often presumed to improve accountability. When agents can see the model family of their counterparts, cross-family trade decreases sharply. The market topology shifts as agents form strong preferences for trading within their own model family, reducing the cross-family exchange through which comparative advantage might operate. This establishes that transparency dramatically alters the trading topology, producing an isolated, family-centric network, without delivering expected improvements in overall task quality or agent wealth.

Modifying agent \emph{disposition} highlights the distinction between individual agent design and market design. Honest and Collaborative instructions do not consistently reduce
settlement friction relative to the neutral baseline. Adversarial
instructions coincide with less cross-family trade. These conditions show that instructions
and population composition can change market behavior, but do not identify
a general effect of honesty, cooperation, or diversity on market surplus.

Finally, the \emph{fierce selection} condition (a compound stress test with six times the baseline replacement rate) severely reduces both output quality and agent wealth (Figure 4). Simply increasing competitive pressure does not optimize the market.

\section{Discussion}
\label{sec:discussion}

\paragraph{Social implications of agent-to-agent markets.}
As AI agents transact on behalf of users and organizations,
agent-to-agent markets become an institutional layer between human
goals and the work delivered. Their rules shape more than technical
efficiency: they influence the reliability of delegated work, the
allocation of tasks and resources, and who bears the cost of failed
evaluation or settlement. Payment friction, identity-based
fragmentation, and quality losses under high turnover can therefore
affect the principals agents represent as well as the platforms
responsible for governing exchange. Agent-market design is consequently
a social-impact problem concerning reliability, accountability, and the
distribution of gains and risks from AI-mediated work.
\paragraph{Toward responsible agent-to-agent markets.}
\diagon\ takes a step toward responsible agent-to-agent markets by
giving researchers a programmable environment for designing and
experimentally evaluating market rules before deployment. The
instantiation studied here demonstrates this use and characterizes one
combination of agents, tasks, and rules. \footnote{Code and additional materials are provided in the supplementary material to support replication and future extensions.} Such evidence can inform platform designers and
policymakers, while field evaluation with platform operators and the
users and organizations represented by agents is an important next
step. Responsible agent markets should ultimately be judged by whether
they deliver reliable and accountable cognitive work to the people and
organizations they serve.

\bibliographystyle{plainnat}
\bibliography{references}

@misc{anthropic2025claudecode,
  title={Claude Code: An Agentic Coding Tool},
  author={{Anthropic}},
  year={2025},
  howpublished={\url{https://github.com/anthropics/claude-code}},
  note={Accessed 2025-03-31}
}

@inproceedings{lin2024collusion,
  title={Strategic collusion of {LLM} agents: Market division in multi-commodity competitions},
  author={Lin, Ryan Y and Ojha, Siddhartha and Cai, Kevin and Chen, Maxwell F},
  booktitle={NeurIPS 2024 Workshop on Language Gamification},
  year={2024}
}

@inproceedings{huang2025gamabench,
  title={Competing Large Language Models in Multi-Agent Gaming Environments},
  author={Huang, Jen-Tse and Li, Eric John and Lam, Man Ho and Liang, Tian and Wang, Wenxuan and Yuan, Youliang and Jiao, Wenxiang and Wang, Xing and Tu, Zhaopeng and Lyu, Michael R.},
  booktitle={The Thirteenth International Conference on Learning Representations},
  year={2025},
  url={https://openreview.net/forum?id=DI4gW8viB6}
}

@article{vaccaro2025negotiations,
  title={Advancing {AI} negotiations: New theory and evidence from a large-scale autonomous negotiation competition},
  author={Vaccaro, Michelle and Caosun, Michael and Ju, Harang and Aral, Sinan and Curhan, Jared R},
  journal={arXiv preprint arXiv:2503.06416},
  year={2025}
}

@article{liu2026agenticpay,
  title={{AgenticPay}: A multi-agent {LLM} negotiation system for buyer--seller transactions},
  author={Liu, Xianyang and Gu, Shangding and Song, Dawn},
  journal={arXiv preprint arXiv:2602.06008},
  year={2026}
}

@inproceedings{liu2025exploring,
  title={Exploring prosocial irrationality for {LLM} agents: A social cognition view},
  author={Liu, Xuan and Zhang, Jie and Shang, HaoYang and Guo, Song and Yang, Chengxu and Zhu, Quanyan},
  booktitle={The Thirteenth International Conference on Learning Representations},
  year={2025},
  url={https://openreview.net/forum?id=u8VOQVzduP}
}

@inproceedings{liu2025cobra,
author = {Liu, Xuan and Shang, HaoYang and Jin, Haojian},
title = {CoBRA: Programming Cognitive Bias in Social Agents Using Classic Social Science Experiments},
year = {2026},
isbn = {9798400722783},
publisher = {Association for Computing Machinery},
address = {New York, NY, USA},
url = {https://doi.org/10.1145/3772318.3790804},
doi = {10.1145/3772318.3790804},
booktitle = {Proceedings of the 2026 CHI Conference on Human Factors in Computing Systems},
articleno = {64},
numpages = {30},
location = {
},
series = {CHI '26}
}

@inproceedings{piatti2024cooperate,
  title={Cooperate or collapse: Emergence of sustainable cooperation in a society of {LLM} agents},
  author={Piatti, Giorgio and Jin, Zhijing and Kleiman-Weiner, Max and Sch{\"o}lkopf, Bernhard and Sachan, Mrinmaya and Mihalcea, Rada},
  booktitle={Advances in Neural Information Processing Systems},
  volume={37},
  year={2024}
}

@article{horton2023llm,
  title={Large language models as simulated economic agents: What can we learn from homo silicus?},
  author={Horton, John J},
  journal={arXiv preprint arXiv:2301.07543},
  year={2023}
}

@inproceedings{li2024econagent,
  title={{EconAgent}: Large language model-empowered agents for simulating macroeconomic activities},
  author={Li, Nian and Gao, Chen and Li, Mingyu and Li, Yong and Liao, Qingmin},
  booktitle={Proceedings of the 62nd Annual Meeting of the Association for Computational Linguistics},
  pages={15523--15536},
  year={2024}
}

@inproceedings{chen2024mechanism,
  title={Mechanism design for large language models},
  author={Duetting, Paul and Mirrokni, Vahab and Paes Leme, Renato and Xu, Haifeng and Zuo, Song},
  booktitle={Proceedings of the ACM Web Conference 2024},
  pages={144--155},
  year={2024}
}

@article{payne2025strategic,
  title={Strategic intelligence in large language models: Evidence from evolutionary game theory},
  author={Payne, Kenneth and Alloui-Cros, Baptiste},
  journal={arXiv preprint arXiv:2507.02618},
  year={2025}
}

@article{hadfield2025economy,
  title={An economy of {AI} agents},
  author={Hadfield, Gillian K and Koh, Andrew},
  journal={arXiv preprint arXiv:2509.01063},
  year={2025}
}

@techreport{shahidi2025coasean,
  title={The {Coasean} Singularity? {Demand}, Supply, and Market Design with {AI} Agents},
  author={Shahidi, Peyman and Rusak, Gili and Manning, Benjamin S and Fradkin, Andrey and Horton, John J},
  year={2025},
  institution={National Bureau of Economic Research}
}

@inproceedings{zhu2025automated,
  title={The automated but risky game: {Modeling} agent-to-agent negotiations and transactions in consumer markets},
  author={Zhu, Shenzhe and Sun, Jiao and Nian, Yi and South, Tobin and Pentland, Alex and Pei, Jiaxin},
  booktitle={ICML 2025 Workshop on Reliable and Responsible Foundation Models},
  year={2025}
}

@article{tomasev2025virtual,
  title={Virtual agent economies},
  author={Tomasev, Nenad and Franklin, Matija and Leibo, Joel Z and Jacobs, Julian and Cunningham, William A and Gabriel, Iason and Osindero, Simon},
  journal={arXiv preprint arXiv:2509.10147},
  year={2025}
}

@misc{openclaw2026,
  title={{OpenClaw}: Your own personal {AI} assistant},
  author={Steinberger, Peter},
  howpublished={\url{https://github.com/openclaw/openclaw}},
  year={2026}
}

@misc{moltbook2026,
  title={Moltbook: The front page of the agent internet},
  author={Schlicht, Matt},
  howpublished={\url{https://www.moltbook.com/}},
  year={2026}
}

@article{goyal2026moltbook,
  title={Social simulacra in the wild: {AI} agent communities on {Moltbook}},
  author={Goyal, Agam and Pal, Olivia and Sundaram, Hari and Chandrasekharan, Eshwar and Saha, Koustuv},
  journal={arXiv preprint arXiv:2603.16128},
  year={2026}
}

@inproceedings{zhuang2023toolqa,
  title={{ToolQA}: A dataset for {LLM} question answering with external tools},
  author={Zhuang, Yuchen and Yu, Yue and Wang, Kuan and Sun, Haotian and Zhang, Chao},
  booktitle={Advances in Neural Information Processing Systems},
  volume={36},
  year={2023}
}

@inproceedings{patil2024bfcl,
  title={The {Berkeley} Function Calling Leaderboard ({BFCL}): From tool use to agentic evaluation of large language models},
  author={Patil, Shishir G and Mao, Huanzhi and Ji, Charlie Cheng-Jie and Yan, Fanjia and Suresh, Vishnu and Stoica, Ion and Gonzalez, Joseph E},
  booktitle={Advances in Neural Information Processing Systems},
  volume={37},
  year={2024}
}

@book{axelrod1984evolution,
  title={The Evolution of Cooperation},
  author={Axelrod, Robert},
  year={1984},
  publisher={Basic Books},
  address={New York}
}

@book{weibull1997evolutionary,
  title={Evolutionary game theory},
  author={Weibull, J{\"o}rgen W},
  year={1997},
  publisher={MIT Press}
}

@inproceedings{hart1987theory,
  title={The theory of contracts},
  author={Hart, Oliver and Holmstr{\"o}m, Bengt},
  booktitle={Advances in economic theory: Fifth world congress},
  volume={1},
  year={1987},
  organization={Cambridge}
}

@article{spence1973job,
  title={Job market signaling},
  author={Spence, Michael},
  journal={The Quarterly Journal of Economics},
  volume={87},
  number={3},
  pages={355--374},
  year={1973},
  publisher={MIT Press}
}

@article{resnick2000reputation,
  title={Reputation systems},
  author={Resnick, Paul and Kuwabara, Ko and Zeckhauser, Richard and Friedman, Eric},
  journal={Communications of the ACM},
  volume={43},
  number={12},
  pages={45--48},
  year={2000},
  publisher={ACM}
}

@book{williamson1985institutions,
  title={The Economic Institutions of Capitalism: Firms, Markets, Relational Contracting},
  author={Williamson, Oliver E},
  year={1985},
  publisher={Free Press},
  address={New York}
}

@book{north1990institutions,
  title={Institutions, Institutional Change and Economic Performance},
  author={North, Douglass C},
  year={1990},
  publisher={Cambridge University Press},
  address={Cambridge}
}

@book{talluri2006theory,
  title={The theory and practice of revenue management},
  author={Talluri, Kalyan T and Van Ryzin, Garrett J},
  volume={68},
  year={2006},
  publisher={Springer Science \& Business Media}
}

@incollection{ricardo2005principles,
  title={From the principles of political economy and taxation},
  author={Ricardo, David},
  booktitle={Readings in the economics of the division of labor: The classical tradition},
  pages={127--130},
  year={2005},
  publisher={World Scientific}
}

@article{tesfatsion2002ace,
  title={Agent-based computational economics: Growing economies from the bottom up},
  author={Tesfatsion, Leigh},
  journal={Artificial Life},
  volume={8},
  number={1},
  pages={55--82},
  year={2002},
  publisher={MIT Press}
}

@incollection{lebaron2006ace,
  title={Agent-based computational finance},
  author={LeBaron, Blake},
  booktitle={Handbook of Computational Economics},
  volume={2},
  pages={1187--1233},
  year={2006},
  publisher={Elsevier}
}

@article{akata2025playing,
  title={Playing repeated games with large language models},
  author={Akata, Elif and Schulz, Lion and Coda-Forno, Julian and Oh, Seong Joon and Bethge, Matthias and Schulz, Eric},
  journal={Nature Human Behaviour},
  volume={9},
  pages={1380--1390},
  year={2025},
  doi={10.1038/s41562-025-02172-y}
}

@article{lin1992self,
  title={Self-improving reactive agents based on reinforcement learning, planning and teaching},
  author={Lin, Long-Ji},
  journal={Machine Learning},
  volume={8},
  number={3--4},
  pages={293--321},
  year={1992},
  publisher={Springer}
}

@article{li2026skillsbench,
  title={SkillsBench: Benchmarking how well agent skills work across diverse tasks},
  author={Li, Xiangyi and Chen, Wenbo and Liu, Yimin and Zheng, Shenghan and Chen, Xiaokun and He, Yifeng and Li, Yubo and You, Bingran and Shen, Haotian and Sun, Jiankai and others},
  journal={arXiv preprint arXiv:2602.12670},
  year={2026}
}

@article{chan2025infrastructure,
  title={Infrastructure for {AI} Agents},
  author={Chan, Alan and Wei, Kevin and Huang, Sihao and Rajkumar, Nitarshan and Perrier, Elija and Lazar, Seth and Hadfield, Gillian K and Anderljung, Markus},
  journal={Transactions on Machine Learning Research},
  year={2025}
}

@inproceedings{kapoor2025advocates,
  title={Position: Build Agent Advocates, Not Platform Agents},
  author={Kapoor, Sayash and Kolt, Noam and Lazar, Seth},
  booktitle={International Conference on Machine Learning},
  year={2025}
}

@inproceedings{jacob2024consensus,
  title={The Consensus Game: Language Model Generation via Equilibrium Search},
  author={Jacob, Athul Paul and Shen, Yikang and Farina, Gabriele and Andreas, Jacob},
  booktitle={International Conference on Learning Representations},
  year={2024}
}

@inproceedings{andreas2022agent,
  title={Language Models as Agent Models},
  author={Andreas, Jacob},
  booktitle={Findings of the Association for Computational Linguistics: EMNLP 2022},
  year={2022}
}

@inproceedings{ye2024eva,
  title={Scalable Reinforcement Post-Training Beyond Static Human Prompts: Evolving Alignment via Asymmetric Self-Play},
  author={Ye, Ziyu and Agarwal, Rishabh and Liu, Tianqi and Joshi, Rishabh and Velury, Sarmishta and Le, Quoc V and Tan, Qijun and Liu, Yuan},
  booktitle={Advances in Neural Information Processing Systems},
  volume={37},
  year={2024}
}

@incollection{akerlof1978market,
  title={The market for “lemons”: Quality uncertainty and the market mechanism},
  author={Akerlof, George A},
  booktitle={Uncertainty in economics},
  pages={235--251},
  year={1978},
  publisher={Elsevier}
}

@article{roth2002economist,
  title={The economist as engineer: Game theory, experimentation, and computation as tools for design economics},
  author={Roth, Alvin E},
  journal={Econometrica},
  volume={70},
  number={4},
  pages={1341--1378},
  year={2002},
  publisher={Wiley Online Library}
}

@article{Richards_Cowser_Nielson_Crandall_2026,
  title={Toward Simulating Networked Societies with Formal Institutions Using AI Agents},
  volume={40},
  url={https://ojs.aaai.org/index.php/AAAI/article/view/41262},
  DOI={10.1609/aaai.v40i46.41262},
  abstractNote={Institutions are key to creating societies that are efficient, fair, and benevolent. Despite their importance, the complexities of human (networked) societies make it difficult to understand how formal institutions form and how they shape human communities. Artificial intelligence (AI) can potentially raise understanding in this regard. Thus, in this paper, we present a simulation model utilizing AI agents to simulate networked societies that contain formal institutions. We then observe the outputs of the resulting model under different societal conditions and formal institutions, and (where applicable) compare and contrast these outputs with political and economic theories. Our model outputs (a) address how inequality impacts societal prosperity, (b) illuminate how institutions can potentially impact poverty, and (c) give insights into the attributes of formal institutions that individuals are inclined to support. These and future simulation models can potentially inform how AI can support the design and development of institutions that facilitate healthier communities and nations.},
  number={46},
  journal={Proceedings of the AAAI Conference on Artificial Intelligence},
  author={Richards, Michael and Cowser, Danny and Nielson, Daniel and Crandall, Jacob W.},
  year={2026},
  month={Mar.},
  pages={39143–39150}
}

@article{Cui_Jiang_Zhou_Qian_Zhang_Wang_2026,
  title={ShortageSim: Simulating Drug Shortages Under Information Asymmetry},
  volume={40},
  url={https://ojs.aaai.org/index.php/AAAI/article/view/41172},
  DOI={10.1609/aaai.v40i45.41172},
  abstractNote={Drug shortages pose critical risks to patient care and healthcare systems worldwide, yet the effectiveness of regulatory interventions remains poorly understood due to information asymmetries in pharmaceutical supply chains. We propose ShortageSim, which addresses this challenge by providing the first simulation framework that evaluates the impact of regulatory interventions on competition dynamics under information asymmetry. Using Large Language Model (LLM)-based agents, the framework models the strategic decisions of drug manufacturers and institutional buyers, in response to shortage alerts given by the regulatory agency. Unlike traditional game theory models that assume perfect rationality and complete information, ShortageSim simulates heterogeneous interpretations on regulatory announcements and the resulting decisions. Experiments on self-processed dataset of historical shortage events show that ShortageSim reduces the resolution lag for production disruption cases by up to 84%, achieving closer alignment to real-world trajectories than the zero-shot baseline. Our framework confirms the effect of regulatory alert in addressing shortages and introduces a new method for understanding competition in multi-stage environments under uncertainty. We open-source ShortageSim and a dataset of 2,925 FDA shortage events, providing a novel framework for future research on policy design and testing in supply chains under information asymmetry.},
  number={45},
  journal={Proceedings of the AAAI Conference on Artificial Intelligence},
  author={Cui, Mingxuan and Jiang, Yilan and Zhou, Duo and Qian, Cheng and Zhang, Yuji and Wang, Qiong},
  year={2026},
  month={Mar.},
  pages={38321–38330}
}

@article{Levy_Segev_Tuisov_Keren_2026,
  title={Multi-Agent Reinforcement Learning for Modeling, Simulating, and Optimizing Energy Markets},
  volume={40},
  url={https://ojs.aaai.org/index.php/AAAI/article/view/41229},
  DOI={10.1609/aaai.v40i45.41229},
  abstractNote={The objective of this study is to advance the optimization of hybrid electricity markets using multi-agent reinforcement learning (MARL). The transition from centralized systems to public–private models introduces significant challenges, including the emergence of independent market players and the increasing integration of renewable energy sources (RESs). These challenges are further intensified by rapidly shifting demand patterns, driven both by energy-intensive data centers and AI inference workloads, as well as by political and societal instabilities. To address these complexities, we develop a formal model of market participants’ behavior and propose a MARL-based framework for optimizing system operator strategies. This framework incorporates dynamic pricing and dispatch scheduling to minimize operational costs, maintain grid stability, and align market incentives. We also present a new, adaptable simulation environment compatible with state-of-the-art MARL methods. Empirical evaluations in increasingly complex scenarios demonstrate the effectiveness of our approach in capturing the dynamic and decentralized nature of modern electricity markets.},
  number={45},
  journal={Proceedings of the AAAI Conference on Artificial Intelligence},
  author={Levy, Matan and Segev, Itay and Tuisov, Alexander and Keren, Sarah},
  year={2026},
  month={Mar.},
  pages={38844–38852}
}

@inproceedings{maevolving,
  title={Evolving Quantitative Reasoning through Self-Play in Digital Twin Markets},
  author={Ma, Tianmi and Huang, Wenxin and Du, Jiawei and Li, Lin and Zhong, Xian and Zhou, Joey Tianyi},
  booktitle={Forty-third International Conference on Machine Learning},
  year={2026}
}

\end{document}